\documentclass[aps,showpacs,floatfix,twocolumn,amsmath,amssymb]{revtex4}
\usepackage{epsfig}  

\begin{document}

\title{Transition magnetic moment of $\Lambda$ $\to$ $\Sigma^0$ in QCD sum rules}
\author{Frank X. Lee and Lai Wang}
\affiliation{Physics Department, The George Washington University,
Washington, DC 20052, USA}

\begin{abstract}
The $\Lambda$ $\to$ $\Sigma^0$ transition magnetic
moment is computed in the QCD sum rules approach. Three independent tensor
structures are derived in the external field method using generalized interpolating 
fields. They are analyzed together with the $\Lambda$ and $\Sigma^0$ mass
sum rules using a Monte-Carlo-based analysis, 
with attention to OPE convergence, ground-state dominance, 
and the role of the transitions in the intermediate states. Relations
between sum rules for magnetic moments of $\Lambda$ and $\Sigma^0$
and sum rules for transition magnetic moment of $\Lambda$ $\to$
$\Sigma^0$ are also examined. Our best prediction for the transition
magnetic moment is $\mu_{\Sigma^0\Lambda}= 1.60\pm 0.07\; \mu_N$. 
A comparison is made with other calculations in the literature.

\end{abstract}
\vspace{1cm}  
\maketitle

\section{Introduction}
\label{intro}

Magnetic moment is an important property in the electromagnetic 
structure of baryons. In the case of octet baryons, 
various theoretical calculations have provided valuable insight into the underlying 
dynamics. The calculations benefited especially from the availability of 
precise experimental information on the magnetic moments.
The $\Lambda$ $\to$ $\Sigma^0$ transition magnetic
moment is considered a member in this octet family,
so any theoretical approach for octet baryon magnetic moments 
should be able to provide an answer for this quantity.
Relative to the attention paid to the individual members of the 
octet baryons, the $\Lambda$ $\to$ $\Sigma^0$ transition magnetic moment 
is less well-known.

The QCD sum rule method is a nonperturbative analytic
formalism firmly entrenched in QCD with minimal modeling~\cite{SVZ79}. In this
method, hadron phenomenology is linked with the interactions
of quarks and gluon through a few parameters:
the QCD vacuum condensates, giving an unique perspective on how the properties
of hadrons arise from nonperturbative interactions in the
QCD vacuum and how QCD works in this context.
The method has been successfully applied to many aspects of strong-interaction physics.
Calculations of the regular baryon magnetic moments have been carried out
in QCD sum rules~\cite{Ioffe84,Chiu86,Pasupathy86,Wilson87,Aliev02,Sinha05,Wang08}.
Limited study of the $\Lambda\to\Sigma^0$ transition magnetic moment
was performed in the traditional QCD sum rule approach~\cite{Zhu98}, 
and in the light-cone formulation of the approach~\cite{Aliev01}.

In this work, we carry out a comprehensive, independent
calculation of the transition magnetic moment of the $\Lambda$
$\to$ $\Sigma^0$ in the external field method in the traditional QCD sum rule approach. 
This would complete the 
picture in the electromagnetic structure of the octet baryons from the 
perspective of QCD sum rules. 
A number of improvements are made. First, we employ generalized interpolating
fields which allow us to use the optimal mixing of interpolating
fields to achieve the best match in a sum rule. Second, we derive a new,
complete set of QCD sum rules at all three tensor structures and
analyze all of them. The previous sum rules, which were mostly
limited to one of the tensor structures, correspond to a special
case of the mixing in our sum rules. Third, we perform a
Monte-Carlo analysis which can give realistic error bars and 
which has become standard nowadays. Fourth,
we analyze the three sum rules using the $\Lambda$ and $\Sigma^0$
mass sum rules derived from the same generalized interpolating fields. 
The performance of each of the sum rules is
examined using the criteria of OPE convergence and ground-state
dominance, along with the role of the transitions in intermediate states.
Finally, we examine the relations between magnetic moment of
$\Lambda$ and $\Sigma^0$ and transition magnetic
moment of $\Lambda$ $\to$ $\Sigma^0$, using the newly-derived QCD sum rules. 

\section{Method}
\label{meth}

The starting point is the time-ordered correlation function in the
QCD vacuum in the presence of a {\em static} background
electromagnetic field $F_{\mu\nu}$:
\begin{equation}
\Pi(p)=i\int d^4x\; e^{ip\cdot x} \langle 0\,|\,
T\{\;\eta^{\Sigma^0}(x)\,
\bar{\eta}^{\Lambda}(0)\;\}\,|\,0\rangle_F. \label{cf2pt}
\end{equation}
On the quark level, it describes a hadron as
quarks and gluons interacting in the QCD vacuum. On the
phenomenological level, it is saturated by a tower of hadronic
intermediate states with the same quantum numbers. By matching the 
two descriptions, a connection can be established between 
hadronic properties and the underlying quark
and gluon degrees of freedom governed by QCD. Here $\eta$ is the
interpolating field that couples to the hadron under consideration. 
The subscript $F$ means that
the correlation function is to be evaluated with an
electromagnetic interaction term added to the QCD Lagrangian:
$
{\cal L}_I = - A_\mu J^\mu,
$
where $A_\mu$ is the external electromagnetic potential and
$J^\mu=e_q \bar{q} \gamma^\mu q$ is the quark electromagnetic current.

Since the external field can be made arbitrarily small, one can
expand the correlation function
\begin{equation}
\Pi(p)=\Pi^{(0)}(p) +\Pi^{(1)}(p)+\cdots,
\end{equation}
where $\Pi^{(0)}(p)$ is the correlation function in the absence of
the field, and gives rise to the mass sum rules of the baryons.
The transition magnetic moment will be extracted from the QCD sum
rules obtained from the linear response function $\Pi^{(1)}(p)$.

The action of the external electromagnetic field is two-fold: it
couples directly to the quarks in the baryon interpolating fields,
and it also polarizes the QCD vacuum. The latter can be described
by introducing new parameters called vacuum susceptibilities.

The interpolating field is constructed from quark fields with the
quantum number of baryon under consideration and it is not unique.
We consider a linear combination of the two standard local
interpolating fields.
\begin{equation}
\begin{array}{l}
\eta^{\Lambda}(uds)={-2\sqrt{\frac{1}{6}}}\epsilon^{abc} [
2(u^{aT}C\gamma_5 d^b)s^c +(u^{aT}C\gamma_5 s^b)d^c
\\
\;\;\;\;\;\;\;\;\;\;\;\;\;\;\;\;\;\;-(d^{aT}C\gamma_5 s^b)u^c
\hspace{2mm}+\beta(2(u^{aT}C d^b)\gamma_5s^c \\\;\;\;\;\;\;\;\;
\;\;\;\;\;\;\;\;\;\;\;+(u^{aT}C s^b)\gamma_5d^c -(d^{aT}C
s^b)\gamma_5u^c )];
\end{array}
\label{infieldL}
\end{equation}
\begin{equation}
\begin{array}{l}
 \eta^{\Sigma^0}(uds)=-\sqrt{2}\epsilon^{abc} [
 (u^{aT}C\gamma_5 s^b)d^c
+(d^{aT}C\gamma_5 s^b)u^c \\\;\;\;\;\;\;\;\;\;
\;\;\;\;\;\;\;\;\;\;+\beta((u^{aT}C s^b)\gamma_5d^c +(d^{aT}C
s^b)\gamma_5u^c) ].
\end{array}
\label{infieldS}
\end{equation}
Here $C$ is the charge conjugation operator; the superscript $T$ means
transpose; and $\epsilon_{abc}$ makes it color-singlet. The
normalization factors are chosen so that correlation functions of
these interpolating fields coincide with each other under
SU(3)-flavor symmetry. The real parameter $\beta$ allows for the
mixture of the two independent currents. The choice advocated by
Ioffe \cite{Ioffe84} and often used in QCD sum rules studies
corresponds to $\beta=-1$. We will take advantage of this freedom
to achieve optimal matching in the sum rule analysis.

\begin{widetext}
\subsection{Phenomenological Representation}
\label{rhs}

We start with the structure of the two-point correlation function
in the presence of the electromagnetic vertex to first order
\begin {equation}
\Pi  (p) = i\int {d^4 x} e^{ipx}  < 0|\eta ^{\Sigma^0} (x)[ -
i\int {d^4 y} A_\mu  (y)J^\mu  (y)]\bar \eta^{\Lambda}(0)|0>_F.
\label{corr}
\end {equation}
Inserting two complete sets of physical intermediate states, we
restrict our attention only to the positive energy ones and write
\begin {equation}
\begin{array}{l}
 \Pi(p) = \int {d^4 x} d^4 y\frac{{d^4 k'}}{{(2\pi )^4 }}\frac{{d^4 k}}{{(2\pi )^4 }}\sum\limits_{N'N} {\sum\limits_{s's} {\frac{{ - i}}{{k'^2  - M_{N'} ^2  - i\varepsilon }}} } \frac{{ - i}}{{k^2  - M_{N}^2  - i\varepsilon }} \\
 e^{ipx} A_{\mu} (y) < 0|\eta ^{\Sigma^0}   (x)|N'k's' >  < N'k's'|J^\mu  (y)| Nks>  <Nks|\bar \eta ^{\Lambda}  (0)|0 > . \\
\end{array}
\end {equation}

The matrix element of the electromagnetic current has the general form
\begin {equation}
\begin{array}{l}
< k's'|J^\mu  (0)| ks>=\bar
u(k',s')[F_1(Q^2)[\gamma_\mu+i{\frac{m_{\Sigma^0}-m_{\Lambda}}{Q^2}}q^{\mu}]-F_2(Q^2)\sigma^{\mu\nu}{\frac{q^\nu}{m_{\Sigma^0}+m_{\Lambda}}}]u(k,s),
 \end{array}
\end {equation}
where $q=k'-k$ is the momentum transfer and $Q^2=-q^2$.
In the following, we treat $\Lambda$ and $\Sigma^0$ mass as degenerate because their
mass difference is small and set $\bar
m=\frac{m_{\Lambda}+m_{\Sigma^0}}{2}$ to be the average mass of
$\Lambda$ and $\Sigma^0$. The Dirac form factors $F_1$ and $F_2$
are related to the Sachs form factors by
\begin {equation}
\begin{array}{l}
G_E(Q^2)=F_1(Q^2)+\frac{Q^2}{(2\bar m)^2}F_2(Q^2)\\
G_M(Q^2)=F_1(Q^2)+F_2(Q^2).
 \end{array}
\end {equation}
At $Q^2=0$, $F_1(0)=1$, $F_2(0)=\mu_{\Sigma^0\Lambda}^a$ which is
the anomalous transition magnetic moment, and
$G_M(0)=F_1(0)+F_2(0)=\mu_{\Sigma^0\Lambda}$ which is the transition magnetic moment.
Note that we consider here the transition $\Lambda\rightarrow \Sigma^0$. 
The moment for the reverse transition is related by a minus sign: 
$\mu_{\Lambda\Sigma^0}=-\mu_{\Sigma^0\Lambda}$.

Writing out explicitly only the contribution of the ground-state
nucleon and denoting the excited state contribution by ESC, we
arrive at
\begin {equation}
\begin{array}{l}
 \Pi (p) = \frac{{i}}{2}\lambda _{\Lambda}\lambda _{\Sigma^0} F_{\mu \nu } [p^2  - \bar m^2  - i\varepsilon ]^{ - 2} (\hat p + \bar m )\{ \frac{{i(\mu_{\Sigma\Lambda}-1)}}{{2\bar m }}\sigma ^{\mu \nu } (\hat p + \bar m ) + i\sigma ^{\mu \nu }  \\
  -( p^\mu  \gamma ^\nu   - p^\nu  \gamma ^\mu  )(\hat p + \bar m )[p^2  - \bar m^2  - i\varepsilon ]^{ - 1} \}  +
  ESC,
 \end{array}
\end {equation}
where $\lambda _{\Lambda}$ and $\lambda _{\Sigma^0}$ are phenomenological parameters
that measure the overlap of the interpolating fields with the ground state.
Examination of its tensor structure using Mathematica reveals that
its has 3 independent combinations, which we name with the following short-hand notation:
\begin{equation}
\mbox{WE}_1 =  F^{\mu \nu } (\hat p\sigma _{\mu \nu } + \sigma _{\mu \nu } \hat p),\;\;
\mbox{WO}_1 =  F^{\mu \nu } \sigma _{\mu \nu },\;\;
\mbox{WO}_2 =  F^{\mu \nu } i(p_\mu \gamma _\nu - p_\nu \gamma.
_\mu )\hat p.
\end{equation}
They are the same as those for the magnetic moment of octet baryons~\cite{Wang08}.
Upon Borel transform the ground state takes the form
\begin {equation}
 \hat B[\Pi (p)] =  - \frac{\lambda _{\Lambda}\lambda _{\Sigma^0}}{{4M^2 }}e ^{ - \bar m^2 /M^2 } 
\{ \frac{1}{{\bar m }}(2\bar m^2 \mu_{\Sigma^0\Lambda} - M^2 (\mu_{\Sigma^0\Lambda}-1) )[WO_1] 
+ \mu_{\Sigma^0\Lambda}[WE_1] 
+ \frac{{2(\mu_{\Sigma^0\Lambda}-1)}}{{\bar m }}[WO_2]\},
\end {equation}
where $M$ is the Borel mass.

Here we must treat the excited states with care. For a generic
invariant function, the pole structure in momentum space can be written as
\begin {equation}
\begin{array}{l}
  \frac{C_{N \leftrightarrow N }^2}{(p^2  -
\bar m^2 )^2}+ \sum\limits_{N^* } {\frac{{C_{N \leftrightarrow N^*
}^2 }}{{(p^2  - \bar m^2 )(p^2  - \bar {m^*}^2 )}}}  +
\sum\limits_{N^* } {\frac{{C_{N^* \leftrightarrow N^* }^2 }}{{(p^2
- \bar {m^*}^2 )^2 }}},
\end{array}
\end {equation}
 where $C_{N \leftrightarrow N}$, $C_{N \leftrightarrow N^*}$ and $C_{N^* \leftrightarrow N^*}$ are constants. The first term is
 the ground state pole which contains the desired transition magnetic
 moment $\mu_{\Sigma^0\Lambda}$. The second term represents the non-diagonal
 transitions between the ground state and the excited states caused
 by the external field. The third term is pure excited state contributions.
 Upon Borel transform, it takes the form
\begin {equation}
\begin{array}{l}
\frac{\lambda _{\Lambda}\lambda
_{\Sigma^0}\mu_{\Sigma^0\Lambda}}{M^2}e^{-\bar m^2/M^2}+e^{-\bar
m^2/M^2}\left[\sum\limits_{N^* } \frac{C_{N\leftrightarrow
N^*}^2}{\bar {m^*}^2-\bar m^2}\left(1-e^{-\left(\bar{m^*}^2-\bar
m^2\right)/M^2}\right)\right]+\sum \limits_{N^* }
\frac{C_{N^*\leftrightarrow N^*}^2}{M^2}e^{-\bar{m^*}^2/M^2}.
\end{array}
\end {equation}
The important point is that the transitions give rise to a
contribution that is not
 exponentially suppressed relative to the ground state. This is a
 general feature of the external-field technique. The strength of
 such transitions at each structure is \emph{a priori} unknown and is an
 additional source of contamination in the determination of the transition magnetic moment
 $\mu_{\Sigma^0\Lambda}$. The standard treatment of the transitions is to approximate the quantity in the square brackets
 by a constant, which is to be extracted from the sum rule along
 with the ground state property of interest. Inclusion of such
 contributions is necessary for the correct extraction of the
 magnetic moments. The pure excited state contributions are
 exponentially suppressed relative to the ground state and can be
 modeled in the usual way by introducing a continuum model and
 threshold parameter.

\subsection{Calculation of the QCD Side}
\label{lhs}

On the quark level, by contracting out the quark pairs in Eq.~(\ref{cf2pt}) 
using the interpolating fields in Eq.~(\ref{infieldL}) and Eq.~(\ref{infieldS}),
we obtain the following master formula in terms of quark propagators,
\begin{eqnarray}
& & \langle\Omega\,|\, T\{\;\eta^{\Sigma^0}(x)\,
\bar{\eta}^{\Lambda}(0)\;\} \,|\,\Omega\rangle
=-2\epsilon^{abc}\epsilon^{a^\prime b^\prime c^\prime} \{\;
\nonumber
\\ & &
 - S_u^{ {aa}'}   \gamma _5   C   S_s^{ {bb}^{\prime ^T}} C\gamma _5   S_d^{ {cc}'}
  +2   S_s^{ {aa}'}   \gamma _5   C   S_u^{ {bb}^{\prime ^T}} C \gamma _5   S_d^{ {cc}'}
 -2    S_s^{ {aa}'}   \gamma _5   C   S_d^{ {bb}^{\prime ^T}} C \gamma _5   S_u^{ {cc}'}
  \nonumber
 \\ & &
  + S_d^{ {aa}'}   \gamma _5   C   S_s^{ {bb}^{\prime ^T}}  C   \gamma _5   S_u^{ {cc}'}
 + S_d^{ {aa}'}     {Tr}[\gamma _5   C   S_s^{ {bb}^{\prime^T}}   C   \gamma _5   S_u^{ {cc}'}]
 -S_u^{ {aa}'}     {Tr}[\gamma _5   C   S_s^{ {bb}^{\prime^T}}   C   \gamma _5   S_d^{ {cc}'}]
  \nonumber
 \\ & &
 -\beta \gamma _5   S_u^{ {aa}'}   \gamma _5   C S_s^{{bb}^{\prime^T}}   C    S_d^{ {cc}'}
 -\beta    S_u^{{aa}'}    C   S_s^{{bb}^{\prime ^T}}  C\gamma _5   S_d^{ {cc}'}   \gamma _5
  +2 \beta \gamma _5   S_s^{ {aa}'}   \gamma _5   C  S_u^{ {bb}^{\prime ^T}}   C    S_d^{ {cc}'}
  \nonumber
 \\ & &
  +2   \beta    S_s^{ {aa}'}    C   S_u^{ {bb}^{\prime ^T}} C \gamma_5   S_d^{ {cc}'}   \gamma _5
   -2   \beta \gamma _5   S_s^{ {aa}'}   \gamma _5   C  S_d^{ {bb}^{\prime ^T}}   C    S_u^{ {cc}'}
  -2   \beta    S_s^{ {aa}'}    C   S_d^{ {bb}^{\prime ^T}} C \gamma_5   S_u^{ {cc}'}   \gamma _5
 \nonumber
 \\ & &
  - \beta ^2 \gamma _5    S_u^{ {aa}'}    C  S_s^{ {bb}^{\prime ^T}}   C    S_d^{ {cc}'}  \gamma _5
 +2    \beta ^2 \gamma _5    S_s^{ {aa}'}    C  S_u^{{bb}^{\prime ^T}}   C    S_d^{ {cc}'}  \gamma _5
 -2    \beta ^2 \gamma _5    S_s^{ {aa}'}    C   S_d^{{bb}^{\prime ^T}}   C    S_u^{ {cc}'}  \gamma _5
 \nonumber
 \\ & &
  +\beta \gamma _5   S_d^{ {aa}'}   \gamma _5   C   S_s^{{bb}^{\prime ^T}}   C    S_u^{ {cc}'}
  +\beta    S_d^{ {aa}'}    C   S_s^{ {bb}^{\prime ^T}}  C   \gamma _5   S_u^{ {cc}'}   \gamma _5
 +\beta \gamma _5   S_d^{ {aa}'}     {Tr}[\gamma _5  C   S_s^{ {bb}^{\prime ^T}}   C    S_u^{ {cc}'}]
  \nonumber
 \\ & &
 +\beta    S_d^{ {aa}'}   \gamma _5    {Tr}[ C  S_s^{ {bb}^{\prime ^T}}   C   \gamma _5   S_u^{ {cc}'}]
 -\beta \gamma _5   S_u^{ {aa}'}     {Tr}[\gamma _5  C   S_s^{ {bb}^{\prime ^T}}   C    S_d^{ {cc}'}]
 -\beta    S_u^{ {aa}'}   \gamma _5    {Tr}[C  S_s^{ {bb}^{\prime ^T}}   C   \gamma _5   S_d^{ {cc}'}]
 \nonumber
 \\ & &
  + \beta ^2 \gamma _5    S_d^{ {aa}'}    C  S_s^{{bb}^{\prime ^T}}   C    S_u^{ {cc}'}  \gamma _5
  + \beta ^2 \gamma _5    S_d^{ {aa}'}   \gamma _5   {Tr}[  C   S_s^{ {bb}^{\prime ^T}}   C    S_u^{ {cc}'}]
  - \beta ^2 \gamma _5    S_u^{ {aa}'}   \gamma _5 {Tr}[ C   S_s^{ {bb}^{\prime ^T}}   C    S_d^{ {cc}'}],
  \label{masterLS}
\end{eqnarray}
where
$
S^{ab}_q (x,0;F) \equiv \langle 0\,|\, T\{\;q^a(x)\,
\bar{q}^b(0)\;\}\,|\,0\rangle_F
$
is the fully interacting quark propagator in the presence of the
electromagnetic field. The propagator can be derived from the 
operator product expansion (OPE). To first order in $F_{\mu\nu}$ and $m_q$
(assume $m_u=m_d=0, m_s\neq 0$), and up to $x^4$, its expression can be found 
in Ref.~\cite{Wang08}.

\subsection{The QCD Sum Rules}
\label{qcdsr}

With the above elements in hand, it is straightforward to evaluate
the correlation function by substituting the quark propagator into
the master formula. We keep terms to first order in the
external field and in the strange quark mass. Terms up to
dimension 8 are considered. The algebra is very tedious. Each
term in the master formula is a product of three copies of the
quark propagator. There are hundreds of such terms summed over various
color permutations. We used a Mathematica package \emph{MathQCDSR}~\cite{Wang11} 
that we developed to carry out the calculations. We confirmed that 
the QCD side has the same tensor structure as the phenomenological side.

Once we have the QCD side (LHS) and the phenomenological side
(RHS), we can construct the sum rules by matching the two sides in the standard way.
Since there are three independent tensor structures, three sum rules can
be constructed. 

At the structure $\mbox{WE}_1$, the  sum rule can be expressed in the following form

\begin{equation}
\begin{array}{l}
 {\rm{c}}_{\rm{1}} L^{-4/9} E_2(w) M^4  + {\rm{c}}_{\rm{2}} m_s\chi a  L^{-26/27}E_1(w) M^2+{\rm{c}}_{\rm{3}} \chi a^2 L^{-4/27}E_0(w)  + {\rm{c}}_{\rm{4}} bL^{-4/9} E_0(w) + ({\rm{c}}_{\rm{5}}  + {\rm{c}}_{\rm{6}} )m_s aL^{-4/9}E_0(w)
 \\+ ({\rm{c}}_{\rm{7}}  + {\rm{c}}_{\rm{8}} )a^2 L^{4/9} \frac{1}{{M^2 }}
  + {\rm{c}}_{\rm{9}} \chi m_0 ^2 a^2 L^{ -18/27} \frac{1}{{M^2 }} + {\rm{c}}_{\rm{10}} m_s m_0 ^2 aL^{ - 26/27} \frac{1}{{M^2
  }}\\=- {\tilde \lambda _{\Lambda}}{\tilde\lambda _{\Sigma^0}}[\frac{{\mu_{\Sigma^0\Lambda} }}{{M^2 }} + A]e^{ - \bar m ^2 /M^2
},
 \end{array}
\label{we1}
\end{equation}

where the coefficients are given by:
\begin{equation}
  \begin{array}{l}
c_{1} = \frac{1}{4} (2 (\beta^2+\beta+1) e_u-(\beta (\beta+2)+2)
e_d)
\\
c_{2} = \frac{1}{4} \beta (3 (\beta+1) e_u-(2 \beta+3) e_d)
\\
c_{3} = -(\frac{1}{6}) (\beta-1) (e_d-e_u) (\beta+f_s+1)
\\
c_{4} = \frac{1}{96} ((e_s+5 e_u) \beta^2+4 e_u \beta-(\beta+2)^2
e_d+4 e_u)
\\
c_{5} = \frac{1}{12} (e_d (-6 f_s+\beta (\beta+3 (\beta+4)
f_s+10)-2)-2 e_u (-3 f_s+\beta \ (\beta+1) (6 f_s+5)-1))
\\
c_{6} = \frac{1}{12} (3 e_s f_s \kappa \phi
\beta^2+(\beta^2+\beta-2) e_d (2 \kappa-\xi)+e_u \ (((\beta-2)
\beta+4) \kappa+(\beta^2+\beta-2) \xi))
\\
c_{7} = \frac{1}{18} (-e_d (\beta^2-2 f_s \beta+5 f_s-1)-(\beta-1)
e_u (5 f_s+\beta (3 \ f_s-1)-1))
\\
c_{8} = \frac{1}{36} (3 e_s f_s \kappa \phi \beta^2+(\beta-1) e_d
((5 f_s+\beta (3 f_s-1)-1) \ \kappa-(\beta+f_s+1)
\xi)\\\;\;\;\;\;\;\;\;+e_u ((\beta^2-2 f_s \beta+5 f_s-1)
\kappa+(\beta-1) \ (\beta+f_s+1) \xi))
\\
c_{9} = \frac{1}{144} (e_s f_s \phi \beta^2+(\beta-1) e_d (5 f_s+2
\beta (f_s+2)+4)-e_u ((f_s+4) \ \beta^2+3 f_s \beta-5 f_s-4))
\\
 c_{10} = \frac{1}{48}((5 e_u (f_s+1)-2 e_s) \beta^2+6 e_u (f_s+1) \beta-(\beta (\beta+6)-12) \
e_d (f_s+1)-12 e_u (f_s+1)).
\end{array}
\label{LSWE1}
\end{equation}
In addition to the standard quark condensate, gluon condensate, and the mixed condensate 
which are rescaled as  
$a=-(2\pi)^2\,\langle\bar{u}u\rangle$, 
$\hspace{2mm} b=\langle g^2_c\, G^2\rangle$, 
$\langle\bar{u}g_c\sigma\cdot G u\rangle=-m_0^2\,\langle\bar{u}u\rangle$.
Three vacuum susceptibilities caused by the external field are introduced:
$\langle\bar{q} \sigma_{\mu\nu} q\rangle_F \equiv
e_q \chi \langle\bar{q}q\rangle F_{\mu\nu}$, 
$\langle\bar{q} g_c G_{\mu\nu} q\rangle_F \equiv
e_q \kappa \langle\bar{q}q\rangle F_{\mu\nu}$, 
$\langle\bar{q} g_c \epsilon_{\mu\nu\rho\lambda} G^{\rho\lambda}
\gamma_5 q\rangle_F \equiv i e_q \xi \langle\bar{q}q\rangle F_{\mu\nu}$.
Note that $\chi$ has the dimension of GeV$^{-2}$, while $\kappa$ and $\xi$ are dimensionless.
The quark charge factors $e_q$ are given in units of electric
charge:
$e_u=2/3, \hspace{4mm} e_d=-1/3, \hspace{4mm} e_s=-1/3$.
Note that we choose to keep the quark charge factors explicit.
The advantage is that it can facilitate the study
of individual quark contribution to the transition magnetic
moment. The parameters $f$ and $\phi$ account for the
flavor-symmetry breaking of the strange quark in the condensates
and susceptibilities:
$
f={ \langle\bar{s}s\rangle \over \langle\bar{u}u\rangle } ={
\langle\bar{s}g_c\sigma\cdot G s\rangle \over
   \langle\bar{u}g_c\sigma\cdot G u\rangle },
\hspace{4mm} \phi={ \chi_s \over \chi }={ \kappa_s \over \kappa
}={ \xi_s \over \xi }.
$
The anomalous dimension corrections of the interpolating fields
and the various operators are taken into account in the leading
logarithmic approximation via the factor
$
L^\gamma=\left[{\alpha_s(\mu^2) \over \alpha_s(M^2)}\right]^\gamma
=\left[{\ln(M^2/\Lambda_{QCD}^2) \over \ln(\mu^2/\Lambda_{QCD}^2)}
\right]^\gamma
$
where $\mu=500$ MeV is the renormalization scale and
$\Lambda_{QCD}$ is the QCD scale parameter. As usual, the pure
excited state contributions are modeled using terms on the OPE
side surviving $M^2\rightarrow \infty$ under the assumption of
duality, and are represented by the factors
$
E_n(w)=1-e^{-w^2/M^2}\sum_n{(w^2/M^2)^n \over n!},
$
where $w$ is an effective continuum threshold and it is in
principle different for different sum rules and we will treat it
as a free parameter in the analysis.
Also, $\tilde \lambda $ is the rescaled current coupling ${\tilde
\lambda \equiv (2\pi)^2\lambda}$. The rescaling is done so there is no explicit factors 
of $\pi$ in the sum rules.

At the structure $\mbox{WO}_1$, the sum rule can be expressed in
the following form
\begin{equation}
\begin{array}{l}
{\rm{c}}_{\rm{1}} m_s L^{-8/9}E_2(w) M^4  + {\rm{c}}_{\rm{2}} \chi
aL^{ - 16/27} E_2(w) M^4  + ({\rm{c}}_{\rm{3}} {\rm{ +
c}}_{\rm{4}} )a E_1(w) M^2  + {\rm{c}}_{\rm{5}} m_s \chi a^2
L^{-16/27}E_0(w) \\+ {\rm{c}}_{\rm{6}} \chi ab L^{-16/27}E_0(w)+
({\rm{c}}_{\rm{7}}+{\rm{c}}_{\rm{8}}) m_s a^2 \frac{1}{{M^2 }}+
{\rm{c}}_{\rm{9}} ab \frac{1}{{M^2 }}
\\= -{\tilde \lambda _{\Lambda}}{\tilde\lambda _{\Sigma^0}} \bar m \left[ \frac{2\mu_{\Sigma^0\Lambda}
}{M^2 } + \frac{\mu_{\Sigma^0\Lambda} -1}{\bar m^2 } +A \right]e^{ -
\bar m ^2 /M^2},
 \end{array}
\label{wo1}
\end{equation}
where the coefficients are
 \begin{equation}
  \begin{array}{l}
  c_{1} = (\beta^2+\beta-2) (e_d-e_u)
\\
c_{2} = \frac{1}{6} ((4 \beta^2+\beta-5) e_u-(\beta^2+\beta-5)
e_d)
\\
c_{3} = \frac{1}{6} (e_d (-3 f_s+\beta (\beta (3
f_s-2)+4)+1)-(\beta-1) e_u (-5 \beta+3 \ (\beta+1) f_s-1))
\\
c_{4} = \frac{1}{24} ((\beta-1) e_u (2 (8 \beta
\kappa+\kappa)+(\beta+8) \xi)+e_d (2 ((7-5 \beta) \ \beta+1)
\kappa+(\beta (2 \beta-7)+8) \xi))
\\
c_{5} = \frac{1}{2} ((\beta^2+\beta+1) e_d-(2 \beta^2+\beta+1)
e_u) (f_s+1)
\\
c_{6} = \frac{1}{144} ((-2 \beta^2+\beta+4) e_d+(\beta-1) (5
\beta+4) e_u)
\\
c_{7} = \frac{1}{18} (e_d ((7 f_s+4) \beta^2+4 f_s \beta+\beta-2
f_s+4)-e_u ((10 f_s+7) \ \beta^2+4 f_s \beta+\beta-2 f_s+4))
\\
c_{8} = \frac{1}{72} (e_d (f_s+\beta (\beta+4 (\beta+1)
f_s+4)+4)-e_u (f_s+\beta (7 f_s \beta+4 \ \beta+4 f_s+4)+4)) (2
\kappa-\xi)
\\
c_{9} = \frac{1}{288} ((-2 \beta^2+\beta+1) e_d-e_u-\beta (\beta
(e_s-e_u)+e_u)).
\end{array}\label{LSWO1}
\end{equation}

At the structure $\mbox{WO}_2$, the sum rule can be expressed in the following form
\begin{equation}
\begin{array}{l}
 {\rm{c}}_{\rm{1}} m_s L^{-8/9} E_1(w) M^2  + {\rm{c}}_{\rm{2}} \chi aL^{ - 16/27} E_1(w) M^2  + ({\rm{c}}_{\rm{3}} {\rm{ + c}}_{\rm{4}} )a E_0(w) + {\rm{c}}_{\rm{5}} m_s \chi a^2 L^{-16/27}E_0(w) + {\rm{c}}_{\rm{6}} m_0 ^2 aL^{-4/9} \frac{1}{{M^2 }} \\
  + c_{\rm{7}} \chi abL^{-16/27} \frac{1}{{M^2 }} + ({\rm{c}}_{\rm{8}}  + {\rm{c}}_{\rm{9}} )m_s a^2  \frac{1}{{M^4 }}{\rm{ + c}}_{{\rm{10}}} ab\frac{1}{{M^4 }} \\
 =   -\frac{{\tilde \lambda _{\Lambda}}{\tilde\lambda _{\Sigma^0}}}{\bar m} [\frac{{2(\mu_{\Sigma^0\Lambda}-1) }}{{M^2 }} + A]e^{ - \bar m ^2 /M^2
},
 \end{array}
\label{wo2}
\end{equation}
where the coefficients for $\Lambda \to \Sigma^0$ at $\mbox{WO}_2$:
 \begin{equation}
  \begin{array}{l}
  c_{1} = \frac{1}{2} (\beta^2+\beta-2) (e_d-e_u)
\\
c_{2} = \frac{1}{6} (\beta-1)^2 (e_d-e_u)
\\
c_{3} = \frac{1}{6} (e_d (\beta^2 (3 f_s+2)-3 (f_s+1))-3
(\beta^2-1) e_u (f_s+1))
\\
c_{4} = \frac{1}{24} (3 (\beta-1) e_u (\beta \xi-2 \kappa)+e_d
((3-2 \beta) \beta \xi-2 \ ((\beta-3) \beta+3) \kappa))
\\
c_{5} = \frac{1}{6} (\beta-1) (e_d-e_u) (-\beta+(\beta+3) f_s+3)
\\
c_{6} = \frac{1}{24} ((-e_s+e_u+4 e_u f_s) \beta^2+3 e_u \beta-e_u
(4 f_s+5)-(\beta-1) e_d (2 \ \beta+4 (\beta+1) f_s+5))
\\
c_{7} = \frac{1}{144} (\beta-1)^2 (e_d-e_u)
\\
c_{8} = \frac{1}{18} (e_u ((10 f_s+7) \beta^2+4 f_s \beta+\beta-2
f_s+4)-e_d ((7 f_s+4) \ \beta^2+4 f_s \beta+\beta-2 f_s+4))
\\
c_{9} = -(\frac{1}{72}) (e_d (f_s+\beta (\beta+4 (\beta+1)
f_s+4)+4)-e_u (f_s+\beta (7 f_s \beta+4 \ \beta+4 f_s+4)+4)) (2
\kappa-\xi)
\\
c_{10} = \frac{1}{288} ((\beta-1) (2 \beta+1) e_d+e_u+\beta (\beta
(e_s-e_u)+e_u)).
\end{array}\label{LSWO2}
\end{equation}

Note that the sum rule from $\mbox{WE}_1$ involves only dimension-even
condensates, so we call this sum rule chiral-even. The sum rule
from both $\mbox{WO}_1$ and $\mbox{WO}_2$ involves only
dimension-odd condensates, so we call them chiral-odd.

\end{widetext}

\section{Sum Rule Analysis}
\label{ana}
The sum rules for transition magnetic moment have the generic form
of OPE - ESC = Pole + Transition, or
\begin{equation}
\Pi_{\Lambda\to\Sigma^0}(QCD,\beta,w,M^2) = {\tilde \lambda
_{\Lambda}}{\tilde\lambda_{\Sigma^0}} \left({\mu_{\Sigma^0\Lambda}
\over M^2} + A\right) e^{-\bar m^2/M^2},
\end{equation}
where $QCD$ represents all the QCD input parameters. The mathematical task 
then boils down to the following: given the function $\Pi_{\Lambda\to\Sigma^0}$ with known
QCD input parameters and the ability to vary $\beta$, find the
phenomenological parameters (mass $\bar{m}$, transition magnetic moment
$\mu_{\Sigma^0\Lambda}$, transition strength $A$, coupling strength
${\tilde \lambda _{\Lambda}}$ and ${\tilde\lambda _{\Sigma^0}}$ ,
and continuum threshold $w$) by matching the two sides over some
region in the Borel mass $M$. A $\chi^2$ minimization is best
suited for this purpose. It turns out that there are too many fit
parameters for this procedure to be successful in general. To
alleviate the situation, we employ the corresponding mass sum
rules for the particles~\cite{Lee02,Wang08}  
which have a similar generic form of OPE - ESC = Pole, or
\begin{equation}
\Pi_{\Lambda}(QCD,\beta,w_1,M^2) = \tilde{\lambda}_\Lambda^2 e^{-m_\Lambda^2/M^2},
\end{equation}
\begin{equation}
\Pi_{\Sigma^0}(QCD,\beta,w_2,M^2) =
\tilde{\lambda}_{\Sigma^0}^2 e^{-m_{\Sigma^2}/M^2}.
\end{equation}
They share some of the common parameters and factors with the
transition sum rules. Note that the continuum thresholds may not
be the same in different sum rules. By taking the following
combination of the transition and mass sum rules,
\begin{equation}
\frac{\Pi_{\Lambda \to \Sigma^0}(QCD,\beta,w,M^2)}{
\sqrt{\Pi_{\Lambda}(QCD,\beta,w_1,M^2) \Pi_{\Sigma^0}(QCD,\beta,w_2,M^2)}} 
\\= \frac{\mu_{\Sigma^0\Lambda}}{M^2}+A,
\label{ratio}
\end{equation}
the couplings $\lambda$ and the exponential factors are canceled
out.
This is the form we are going to implement. By plotting the two
sides as a function of $1/M^2$, the slope will be the transition
magnetic moment $\mu_{\Sigma^0\Lambda}$ and the intercept the
transition strength$A$. The linearity (or deviation from it) of
the left-hand side gives an indication of OPE convergence and the
role of excited states. The two sides are expected to match for a
good sum rule. This way of matching the sum rules has two
advantages. First, the slope, which is the transition magnetic
moment of interest, is usually better determined than the
intercept. Second, by allowing the possibility of different
continuum thresholds, we ensure that both sum rules stay in their
valid regimes.

We use the Monte-Carlo procedure~\cite{Derek96,Wang08,Wang09} 
to carry out the search.  In this method, the entire phase-space of
the input QCD parameters is explored simultaneously, and is mapped
into uncertainties in the phenomenological parameters. This leads
to more realistic uncertainty estimates than traditional approaches.

The QCD input parameters are given as follows. The condensates are
taken as $a=0.52\; GeV^3$, $ b=1.2\; GeV^4$, $ m^2_0=0.72\; GeV^2$.
For the factorization violation parameter, we use $\kappa_v=2.0$.
The QCD scale parameter is restricted to $\Lambda_{QCD}=0.15$ GeV.
The vacuum susceptibilities have been estimated in studies of
nucleon magnetic moments~\cite{Ioffe84,Chiu86,Lee98b,MAM}, but the
values vary depending on the method used. We use $\chi=-6.0\;
GeV^{-2}$ and $\kappa=0.75$, $\xi=-1.5.$
Note that $\chi$ is almost an order of magnitude larger than
$\kappa$ and $\xi$, and is the most important of the three. The
strange quark parameters are placed at $ m_s=0.15\; GeV$,
$f=0.83$, $\phi=0.60$~\cite{Pasupathy86,Lee98b}. These input
parameters are just central values. We will explore sensitivity to
these parameters by assigning uncertainties to them in the
Monte-Carlo analysis.
%

\section{Results and Discussion}
\label{res}

\subsection{Transition magnetic moments of $\Lambda\to\Sigma^0$}

We analyzed all three sum rules. For each
sum rule, we have in principle 6 parameters to determine:
$\mu_{\Sigma^0\Lambda}$, $A$, $w$, $w_1$, $w_2$, $\beta$. But a
search treating all six parameters as free does not work because
there is not enough information in the OPE to resolve them. 
Here, we use the freedom to vary $\beta$ as an advantage to find the optimal
match. Another parameter that can be used to our advantage is the
continuum threshold $w_1$ for the mass sum rule. We fix it to the
value that gives the best solution to the mass sum rule
independently. For $\Lambda$, $w_1=1.60$ GeV; for $\Sigma^0$,
$w_2=1.66$ GeV. This way the transition magnetic moment sum rule
and the mass sum rule can stay in their respective valid Borel
regimes. This leaves us with three parameters:
$\mu_{\Sigma^0\Lambda}$, $A$, $w$. Unfortunately, a three-parameter
search is either unstable or returns values for $w$ smaller than
the particle mass, an unphysical situation. Again we think this is
a symptom of insufficient information in the OPE. So we are forced
to fix the continuum threshold $w$ that corresponds to the best
match for the central values of the QCD parameters.

The Borel window is determined by the following two criteria: OPE
convergence and ground-state dominance. It is done iteratively.
For each value of $\beta$, we adjust the Borel window until the
best solution is found. Based on the quality of the match, the
broadness of the Borel window and its reach into the lower end (non-perturbative regime),
the size of the continuum contribution, and the OPE convergence,
we find that the chiral-odd sum rule at structure $\mbox{WO}_2$ is
more reliable than the chiral-even sum rule at structure
$\mbox{WE}_1$ and chiral-odd sum rule at structure $\mbox{WO}_1$.
The physical reason is that the contributions of positive- and
negative-parity excited states partially cancel each other in the
chiral-odd sum rules whereas they add up in the chiral-even sum
rules. Since the $\mbox{WO}_2$ sum rule has power corrections up
to $1/{M^4}$ while $\mbox{WO}_1$ sum rule power corrections only
goes up to $1/{M^2}$, the $\mbox{WO}_2$ sum rule is expected
to be more reliable than the $\mbox{WO}_1$ sum rule.
In the following, we present some detailed analysis of this sum rule 
in order to demonstrate how it produces the results. 

\begin{figure}[htbp]
\centerline{\psfig{file=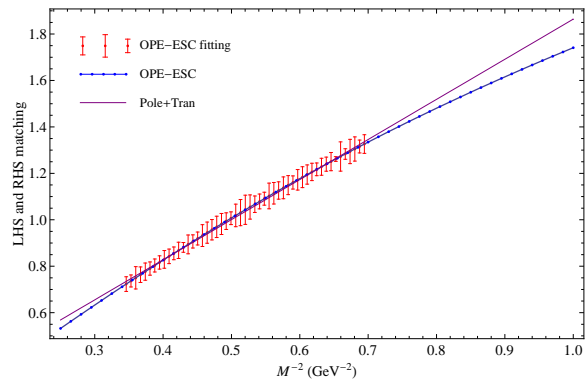,width=8.0cm}}
\vspace*{-0.20cm} \caption{The matching for the chiral-odd QCD
sum rule at structure $\mbox{WO}_2$ in Eq.~(\protect\ref{wo2}) 
as prescribed in Eq.~(\protect\ref{ratio}).
The error band is derived from about 2000 QCD parameter sets in the Monte Carlo analysis.}
\label{massmatch1}
\end{figure}

Fig.~\ref{massmatch1} shows the matching
of the two sides in Eq.~(\protect\ref{ratio}). The left hand side
is normalized OPE minus ESC contribution and the right hand side
is pole plus transition contribution. According to the right-hand
side of this equation, the plot should be linear as a function of
$1/M^2$, so we try to match the left hand side over a Borel window
by a linear line. We can see that the two curves agree very well
over the Borel window 1.2 to 1.7. Based on this match, the slope
$1.60$ gives the transition magnetic moment $\mu_{\Sigma^0\Lambda}$,
and the intercept $0.17$ gives the transition contribution $A$ between intermediate states.
This match was performed at $\beta = -0.2$ and $w = 1.70$ GeV.

\begin{figure}[htbp]
\centerline{\psfig{file=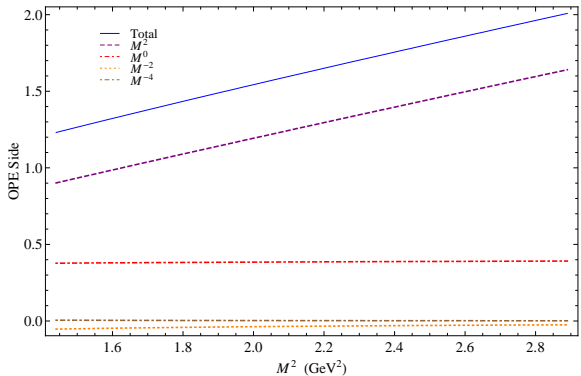,width=8.0cm}}
\vspace*{-0.20cm} \caption{Contributions from individual terms in the OPE for
the $\mbox{WO}_2$ sum rule.}\label{massope2}
\end{figure}

Figure~\ref{massope2} shows how the various terms in the OPE
contribute to the determination of the transition magnetic moment. The
$M^2$ term, which contains the contributions from the condensates
$\chi a$ and strange quark mass $m_s$, is the dominant term 
over the whole Borel region.
The next leading contribution comes from $M^0$
term which involves condensates $a$ and $m_s\chi a^2$, and is
constant over the Borel region. The $M^{-2}$ term is from
condensates $m_0^2 a$ and $\chi a b$, and is slightly negative over
the region. The $M^{-4}$ term comes from
condensates $m_s a^2$ and $ab$, and is almost zero. These results 
confirm that the $WO_2$ sum rule has good OPE convergence 
and why it is the best sum rule out of the three structures.
It also shows the importance of the vacuum susceptibility $\chi$ in the transition.

\begin{figure}[htbp]
\centerline{\psfig{file=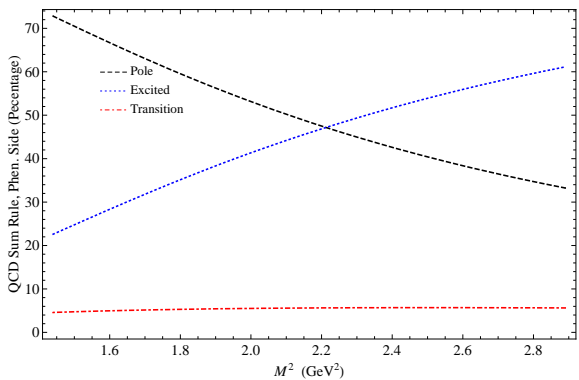,width=8.0cm}}
\vspace*{-0.20cm} \caption{The three terms (pole, transition, and ESC) 
as a percentage of the  spectral representation 
for the $\mbox{WO}_2$ sum rule.}\label{massphen2}
\end{figure}

Figure~\ref{massphen2} show various contributions in the phenomenological 
representation in $\mbox{WO}_2$ as a percentage of the total.
We need to make sure that the ground-state pole is dominant. 
We can see that it is about 70\% at the lower end of the Borel window. The
transition contribution is small in this sum rule. It is
consistently smaller than the excited-state contribution and has a
weak dependence on the Borel mass. The excited-state contribution
starts small, then grows with $M^2$ as expected from the continuum
model. We cut off the Borel window when excited-state contribution
goes too big, here it is around 60\%.

\begin{figure}[htbp]
\centerline{\psfig{file=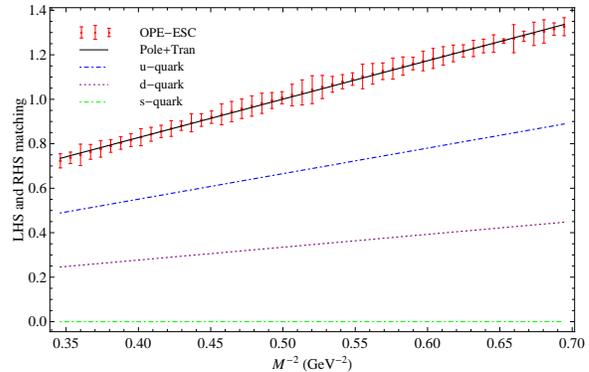,width=8.0cm}}
\vspace*{-0.20cm} \caption{Individual quark contributions 
in the $\mbox{WO}_2$ sum rule.  } \label{massmatch2}
\end{figure}

To gain a deeper understanding of the underlying quark-gluon dynamics, it is useful to
consider the individual quark sector contributions to the
transition magnetic moment. This can be done by dialing the 
corresponding charge factors $e_q$ in the sum rules which were explicitly 
kept for this purpose. 
Figure~\ref{massmatch2} shows a detailed matching
of the two sides on Eq.~(\protect\ref{ratio}) for the $\mbox{WO}_2$ sum rule,
with individual contributions from u-, d- and s-quark. The four slopes are $0.80$
for total $\mu_{\Sigma^0\Lambda}$ and $0.53$, $0.27$ and $0$ for u-,
d- and s-quark. We see that the u-quark contributes about 67\% to the
total transition magnetic moment, d-quark about 33\%,
 and s-quark almost zero. The slope for
$\mbox{WO}_2$ comes from $\frac{2(\mu_{\Sigma^0\Lambda}-1)}{\bar
m}$, so it gives $\mu_{\Sigma^0\Lambda}=1.60\; \mu_N$. To get the
effective individual quark transition magnetic moment, we can use
the percentage of individual contribution from the total
transition magnetic moment. Then we have
$\mu_{\Sigma^0\Lambda}^u=1.05\;\mu_N$,
$\mu_{\Sigma^0\Lambda}^d=0.53\;\mu_N$ and $\mu_{\Sigma^0\Lambda}^s \approx 0\;
\mu_N$.
A closer examination reveals that the smallness of the s-quark contribution 
is due to an almost exact cancellation of the terms involving the s-quark, 
even though these terms are not small themselves.
%
\begin{table*}[thb] 
\caption{Results for $\Lambda \to \Sigma^0$ from the three QCD sum rules.
The ten columns correspond to, from left to right: structure,
$\beta$ value, Borel region in which the two sides of the QCD sum
rule are matched, continuum threshold, percentage of excited state
contribution on the phenomenological representation, transition strength, extracted
transition magnetic moment in unit of nuclear magnetons, 
and effective u-, d- and s-quark
transition magnetic moment. The errors are derived from 2000
samples in the Monte-Carlo analysis with 10\% uncertainty on all
QCD input parameters.} \label{tabwe1}
 \begin{tabular*}{0.75\textwidth}{@{\extracolsep{\fill}}lcccccccccc}
\hline\hline
Structure   & $\beta$  & Region   & $w$   &ESC  & A  &$\mu_{\scriptscriptstyle \Sigma^0\Lambda}$  &$\mu_{\Sigma^0\Lambda}^u$&$\mu_{\Sigma^0\Lambda}^d$&$\mu_{\Sigma^0\Lambda}^s$\\
                      &          & (GeV)         & (GeV)     & (\%)    & (GeV$^{-2}$)     & $(\mu_{\scriptscriptstyle N})$&$(\mu_{\scriptscriptstyle N})$&$(\mu_{\scriptscriptstyle N})$&$(\mu_{\scriptscriptstyle N})$\\ \hline
WE1 & -0.2 & 1.30 to 1.80 & 1.80 &40-80& 0.11(1)  & 1.57(8)  & 1.04(8)&0.53(3)&0(0)\\
WO1 & -0.2 & 1.40 to 2.00 & 1.70 &70-80& -0.46(9) & 1.63(11)  & 1.08(11)&0.55(4)&0(0)\\
WO2 & -0.2 & 1.20 to 1.70 & 1.70 &  20-60& 0.17(4)   & 1.60(7) & 1.05(10) & 0.53(3)& 0(0)\\
\hline \hline
\end{tabular*}
\end{table*}
%

We analyzed sum rules at structures $\mbox{WE}_1$ and $\mbox{WO}_1$
in a similar fashion. The full results determined at the structures are
given in Table~\ref{tabwe1}. We see that our calculated
transition magnetic moment is $1.60\pm 0.07\;\mu_N$ from the most
reliable sum rule $\mbox{WO}_2$. The other two sum rules also give
close results around $1.60\;\mu_N$. 
Our errors are derived from Monte-Carlo distributions which give
more realistic estimation of the uncertainties. We find about
10\% accuracy for the transition magnetic moment in our
Monte-Carlo analysis, resulting from 10\% uniform uncertainty in
all the QCD input parameters. The uncertainties in the QCD
parameters can be non-uniform. For example, we tried the
uncertainty assignments (which are quite conservative) in
Ref.~\cite{Derek96}, and found about 30\% uncertainties in our
output. One choice for $\beta$ here is $-0.2$ because it minimizes the
perturbative term in the mass sum rule~\cite{narison}. The excited
state contribution at structure $\mbox{WO}_2$ is 20-60\%, which
gives reasonable pole dominance over the whole Borel region.
On the other hand, the  sum rules at $\mbox{WE}_1$ and $\mbox{WO}_1$
structures have larger contributions from the excited states
over the Borel region. For this reason, we say the sum rules from
$\mbox{WE}_1$ and $\mbox{WO}_1$ are less reliable. Another measure 
of reliability is the parameter $A$, which is contamination from 
transitions between intermediate states. 
The A for the $\mbox{WO}_1$ sum rule is larger than those for the 
$\mbox{WE}_1$ and $\mbox{WO}_2$ sum rules, making it less reliable.

\subsection{Relations between QCD sum rules}
It was pointed in Ref.~\cite{Ozpineci03} that it is possible to 
derive QCD rum rules for $\Lambda\to\Sigma^0$ from 
those for $\Sigma^0$ and $\Lambda$. Here we would like to examine the issue 
using the generalized interpolating fields for $\Lambda$ and $\Sigma^0$ in
Eq.~(\ref{infieldL}) and Eq.~(\ref{infieldS}). 

The two-point correlation functions for magnetic moments of
$\Lambda$ and $\Sigma^0$ are constructed from time-ordered product of operators
$T\{\eta^{\Lambda}(x)\, \bar{\eta}^{\Lambda}(0)\}$ and
$T\{\eta^{\Sigma^0}(x)\, \bar{\eta}^{\Sigma^0}(0)\}$. 
The result is a master formula in terms of fully-interacting quark propagators 
in the QCD vacuum.
Using the master formulas in Ref.~\cite{Wang08}, 
we have checked the following relations between the correlation functions
\begin{eqnarray}
2[\tilde\Pi^{\Sigma^{0}}_{d\leftrightarrow s}+
\tilde\Pi^{\Sigma^{0}}_{u\leftrightarrow s}]-
\Pi^{\Sigma^{0}}=3\Pi^{\Lambda},
\\
2[\tilde\Pi^{\Lambda}_{d\leftrightarrow s}+
\tilde\Pi^{\Lambda}_{u\leftrightarrow s}]-
\Pi^{\Lambda}=3\Pi^{\Sigma^{0}}. \label{crossLa}
\end{eqnarray}
Here $\tilde\eta^{\Lambda}_{d\leftrightarrow s}$ means the
correlation function of $\Lambda$ but with d-quark and s-quark interchanged.
The relations imply that one can obtain the QCD sum rules for $\Sigma^0$ 
by simple substitutions from the QCD sum rules for $\Lambda$ and vice versa.
We confirmed that relations not only at the correlation function level, 
but at the level of the derived sum rules in ~\cite{Wang08}.
It serves as an independent check of the QCD sum rules in Ref.~\cite{Wang08}.

The same argument can be extended to the $\Lambda\to\Sigma^0$ transition moment.
There exist the following relations
\begin{eqnarray}
\tilde\Pi^{\Sigma^{0}(u\leftrightarrow s)}-
\tilde\Pi^{\Sigma^{0}(d\leftrightarrow s)} =\sqrt{3} \Pi^{\Sigma^{0}\Lambda},
 \\
\tilde\Pi^{\Lambda(u\leftrightarrow s)}-
\tilde\Pi^{\Lambda(d\leftrightarrow s)} =-\sqrt{3}\Pi^{\Sigma^{0}\Lambda}.
\label{crossSi}
\end{eqnarray}
This suggests that one can derive the QCD sum rules for the 
$\Lambda\to\Sigma^0$ transition magnetic moment by simple substitutions, 
starting from those for either $\Sigma^0$ or $\Lambda$.
In fact, we have calculated the QCD sum rules for $\Lambda\to\Sigma^0$ separately, 
and used the relations as independent checks of our calculations in this work 
and those in Ref.~\cite{Wang08}. 
This was done both at the level of the master formula in Eq.~(\ref{masterLS}) and 
at the level of the coefficients in the final QCD sum rules.
The calculations has also served as a check of the Mathematica package~\cite{Wang11}.

\subsection{Comparison of results}
Finally, we compare in Table~\ref{tabcom} our result with experiment and other calculations.
The list is only a representative sample, and necessarily incomplete.
Unlike the other members of the octet, there was only one measurement 
for the $\Lambda \to \Sigma^0$ transition magnetic moment. It was done at Fermilab 
by utilizing the Primakoff effect~\cite{Petersen86}.
Our result agrees with experiment.
In the simple quark model, the transition moment can be related to the magnetic 
moment of $\Sigma^+$ and $\Sigma^-$ by 
$\mu_{\Sigma^0\Lambda}=(\mu_{\Sigma^+} - \mu_{\Sigma^-})\sqrt{3}/4$. 
In terms of effective quark moments it is given by
$\mu_{\Sigma^0\Lambda}=(\mu_u - \mu_d)/\sqrt{3}$. Using the values 
$\mu_u=1.852 \;\mu_N$ and $\mu_d=-0.972 \;\mu_N$, 
it means that the u-quark contribution is about 67\% 
of the total, the d-quark 33\%, and the s-quark zero. 
Remarkably, our QCD sum rule results from the study of 
individual quark contributions are consistent with this composition.
The lattice QCD result in Ref.~\cite{Derek91} is quoted here by 
a scale factor of 1.18 as suggested in Ref.~\cite{Derek96a}.

\begin{table}
\caption{Comparison of results for the 
$\Lambda \to \Sigma^0$ transition magnetic moments in nuclear magnetons.}
\label{tabcom}
\begin{tabular}{ll}
\hline\hline
Method   & $\mu_{\Sigma^0\Lambda}$ \\
\hline
Experiment~\protect\cite{pdg10} &  1.61(8) \\
QCD sum rule (this work) &  1.60(7)  \\
QCD sum rule~\protect\cite{Zhu98} &  1.5 \\
Light-cone QCD sum rule~\protect\cite{Aliev01} &  1.60 \\
Simple quark model~\protect\cite{pdg10,Franklin84} & 1.57(1) \\
Relativized quark model~\protect\cite{Warns91} & 1.04 \\
Quark model fits~\protect\cite{Bohm82} & 1.81 \\
Chiral quark model~\protect\cite{Dahiya03} & 1.71, 1.66, 1.61 \\
Lattice QCD~\protect\cite{Derek91} &  1.36 \\
Chiral Perturbation Theory~\protect\cite{Ulf97,Ulf01} & 1.42(1), 1.61(1) \\
Large-N ChPT~\protect\cite{Gio11} & 1.576 \\
\hline \hline
\end{tabular}
\end{table}
%

\section{Conclusion}
\label{con}

We have carried out a comprehensive study of the $\Lambda$ $\to$ $\Sigma^0$
transition magnetic moment using the method of QCD sum rules. 
We derived a complete set of QCD sum rules using generalized
interpolating fields and analyzed them by a Monte-Carlo analysis.
We determined the transition magnetic moment from the slope of
straight lines. The linearity displayed from the OPE side matches
almost perfectly with the phenomenological side in all cases. Out
of the three independent structures, we find that the sum rule
from the $\mbox{WO}_2$ structure is the most reliable based on OPE
convergence and ground-state pole dominance. Our best prediction for
the transition magnetic moment is $\mu_{\Sigma^0\Lambda}=1.60\pm
0.07\;\mu_N$. More detailed results are listed in
Table~\ref{tabwe1}. The extracted transition magnetic moment is 
in good agreement experiment. Our Monte-Carlo analysis reveals 
that there is an uncertainty on the
level of 10\% in the transition magnetic moment if we assign 10\%
uncertainty in the QCD input parameters. We find that for all
three sum rules, the u-quark contributes about $2/3$ of the total
transition magnetic moment, d-quark about $1/3$, 
while s-quark negligible due to an almost exact cancellation. This result 
agrees with the simple quark model.
We verified certain relations that exist between the QCD sum rules 
and used them to check our calculations.
Contrasted with results from a variety of other calculations,
this study adds an unique perspective on the inner workings of the 
$\Lambda$ $\to$ $\Sigma^0$ transition magnetic moment 
in terms of the non-perturbative nature of the QCD vacuum.

\begin{acknowledgments}
This work is supported in part by U.S. Department of Energy under grant
DE-FG02-95ER-40907.
\end{acknowledgments}


\end{document}